\documentclass[%
 reprint,
 amsmath,amssymb,
 aps,
 floatfix,
]{revtex4-2}

\pdfoutput=1 

\usepackage{graphicx}
\usepackage{dcolumn}
\usepackage{algorithm}
\usepackage{algpseudocode}
\usepackage{booktabs}
\usepackage{multirow}
\usepackage{bm}
\usepackage{aas_macros}
\usepackage{fontawesome}
\usepackage{hyperref}
\usepackage[dvipsnames]{xcolor}
\usepackage{physics}

\newcommand\myshade{80}
\colorlet{mylinkcolor}{ForestGreen}
\colorlet{mycitecolor}{Red}
\colorlet{myurlcolor}{violet}
\hypersetup{
  linkcolor  = mylinkcolor!\myshade!black,
  citecolor  = mycitecolor!\myshade!black,
  urlcolor   = myurlcolor!\myshade!black,
  colorlinks = true
}

\newcommand{\FigRef}[1]{Fig.~(\ref{#1})}
\newcommand{\EqnRef}[1]{Eqn.~(\ref{#1})}
\newcommand{\TabRef}[1]{Tab.~(\ref{#1})}
\newcommand{\AlgRef}[1]{Alg.~(\ref{#1})}

\newcommand{\software}[1]{#1}
\newcommand{\asusy}{a^{\mathrm{SUSY}}_{\mu}}
\newcommand{\rdchi}{\Omega_{\chi}h^{2}}

\begin{document}


\title{\texorpdfstring{Simulation Based Inference for Efficient Theory Space Sampling:\\ an Application to Supersymmetric Explanations of the Anomalous Muon (g-2)}{Simulation Based Inference for Efficient Theory Space Sampling: an Application to Supersymmetric Explanations of the Anomalous Muon (g-2)}}

\author{Logan Morrison}
\author{Stefano Profumo}
\author{Nolan Smyth}
\author{John Tamanas}
\email{jtamanas@ucsc.edu}
\affiliation{Department of Physics and Santa Cruz Institute for Particle Physics
University of California, Santa Cruz, CA 95064, USA
}

\date{\today}

\begin{abstract}
    For the purpose of minimizing the number of sample model evaluations, we propose and study algorithms that utilize (sequential) versions of likelihood-to-evidence ratio neural estimation.
    We apply our algorithms to a supersymmetric interpretation of the anomalous muon magnetic dipole moment in the context of a phenomenological minimal supersymmetric extension of the standard model, and recover non-trivial models in an experimentally-constrained theory space.
    Finally we summarize further potential possible uses of these algorithms in future studies. \href{https://www.github.com/jtamanas/MSSM}{\faGithub}
  \end{abstract}

\maketitle

\section{Introduction}


The pursuit of understanding beyond the Standard Model (BSM) physics theories in the context of experimental results is the cornerstone of much research in high energy physics (HEP). However, despite the increasing accuracy of weak-scale observables, it remains non-trivial to perform the inverse calculation, i.e., to determine regions of parameter space yielding the observables.


Mapping out regions of parameter space that produce given observables is especially difficult for theories with many free parameters -- a famous example being the minimal supersymmetric model (MSSM) (see, e.g., \cite{Chung:2003fi} for a review) and down-sized versions thereof, in parameter space, such as the phenomenological MSSM (pMSSM) \cite{Berger_2009, AbdusSalam:2009qd}. The exploration of the parameter space of these theories is typically carried out, in the HEP literature, by defining a search space in the space of pMSSM parameters, uniformly sampling it, and then rejecting all samples which do not conform with theoretical or experimental constraints (see, e.g., \cite{Roszkowski:2014iqa, Bertone_2016, VanBeekveld:2021tgn}, albeit there is a very vast literature on the topic). However, the number of required samples grows exponentially with the dimension of the search space, so these methods are extremely computationally intensive, if not intractable, when calculations of the output predictions for a given set of input parameters are slow.

There has been a recent upsurge of developments in the broader scientific literature regarding the inverse problem of restricting parameter spaces of a forward model purely through sampling (i.e. calculating observables from model parameters; see e.g. \cite{pydelfi, swyft, cole} for a few recent examples).  The introduction of neural networks to simulation-based inference (SBI) frameworks, in particular, has lead to explosive improvements in accuracy and precision and ablation studies comparing them all \cite{snl, snp, snre, sbi-benchmark}. 


 In this work, we introduce the sequential neural ratio estimation (SNRE) algorithm to the problem of sampling from an experimentally constrained version of the pMSSM. Although not explicitly stated, recent work \cite{hollingsworth} has used methods from the simulation-based inference literature -- namely neural likelihood estimation. We show how that method fits into the larger SBI framework and discuss an alternative, though related, approach that uses likelihood-to evidence ratio neural estimation. This fits in the context of ``likelihood-free" methods in which drawing samples from a forward model is possible, but the likelihood is expensive or intractable to evaluate. Additionally, we demonstrate sequential versions of two algorithms that can significantly reduce the number of model evaluations required. 
 

Specifically, here we are interested in the following example case study: we perform sampling of the pMSSM parameter space, producing experimentally viable predictions for (i) the relic density of dark matter ($\Omega_\chi h^2$) in the form of the lightest neutralino, which has mass $M_\chi$, (ii) the Higgs mass ($m_h$), (iii) the anomalous magnetic moment of the muon ($a_{\mu}$), and (iv) WIMP-nucleon cross sections ($\sigma^{\textrm{SI}}$).
 

We organize the manuscript as follows: \S~\ref{sec:sbi} details the general SBI framework and describes the SNRE algorithm in particular; \S~\ref{sec:pmssm} details the search space of the pMSSM to which we will be applying SNRE; in  \S~\ref{sec:benefits} we demonstrate the improved sampling efficiency of sequential algorithms; we show an application of SNRE to an experimentally constrained pMSSM parameter space in \S~\ref{sec:sbi-pmssm}; and finally, in \S~\ref{sec:conclusion} we summarize our findings and further discuss potential future uses of SBI. 

\section{Simulation-Based Inference}
\label{sec:sbi}


This work aims to use the simulation-based inference (SBI) framework to sample from a large-dimensional parameter space while evaluating the model output as few times as possible for computational efficiency. There are essentially four overarching approaches to SBI: rejection sampling with approximate Bayesian computation (ABC), posterior estimation (NPE), likelihood-to-evidence ratio estimation (NRE), and likelihood estimation (NLE). For a full review of simulation-based inference and other methods therein, we refer the Reader to Ref.~\cite{sbi-review}.

We begin by stating Bayes' rule:
\begin{equation}
    \label{eq:bayes}
    P(\theta | X) = \frac{P(X | \theta) P(\theta)}{P(X)}
\end{equation}
where $P(X | \theta)$ is the likelihood of the data given the model parameters, $P(\theta)$ is the prior probability of the model parameters, and $P(X)$ is the evidence. In this work, $\theta$ refers to supersymmetric parameters, and $X$ refers to various observables of our choosing, e.g., the Higgs mass, the DM relic density, and the muon anomalous magnetic moment.

The NRE approach begins by approximating the likelihood-to-evidence ratio, $r(X, \theta) = \frac{P(X | \theta)}{P(X)}$, and then using it to sample from the posterior distribution of the model parameters with Hamiltonian Monte Carlo (HMC). HMC can be thought of as a Markov Chain Monte Carlo (MCMC) \cite{metropolis, hastings} algorithm where a chain's position and momentum are sampled from a joint distribution and then evolved according to Hamilton's equations. HMC makes use of derivatives to calculate its transitions, which makes it not directly applicable using the \software{micrOMEGAs} \cite{micromegas} package, but it can still be used on the neural network side. For a full review of HMC, see \cite{hmc}.

\subsection{Likelihood-to-Evidence Ratio Estimation}

Here, we will  follow the approach of Ref.~\cite{snre} to approximate the likelihood-to-evidence ratio: Assume we have two classes of simulations, $Y_0$ and $Y_1$, each of which is composed of pairs of $\{X, \theta \}$. We define $Y_0$ and $Y_1$ such that $Y_0 = \{X, \theta\} \sim P(X)P(\theta)$ and $Y_1 = \{X, \theta\} \sim P(X | \theta)$. We use curly brackets to denote a set of values. The optimal classifier (one which minimizes the binary cross-entropy loss) between $Y_0$ and $Y_1$ is given by 
\begin{equation}
    d^{*}(X, \theta) = \frac{P(X|\theta)}{P(X|\theta) + P(X)P(\theta)}
\end{equation}
which can, in turn, be used to express the likelihood-to-evidence ratio:
\begin{equation}
    r^{*}(X, \theta) = \frac{P(X|\theta)}{P(X)} = \frac{P(X,\theta)}{P(X)P(\theta)} = \frac{d^{*}(X, \theta)}{1 - d^{*}(X, \theta)},
\end{equation}
where we used the fact that the joint distribution $P(X,\theta) = P(X|\theta)P(\theta)$. Thus, approximating $r^{*}(X, \theta)$ is equivalent to finding the optimal classifier $d^{*}(X, \theta)$, which we can accomplish by training a classifier to distinguish between $Y_0$ and $Y_1$. For the rest of this work, we shall use $r(X, \theta)$ to denote the approximate likelihood-to-evidence ratio.

In practice, we produce samples of $Y_1$ by first drawing $\theta$ from the prior, $P(\theta)$, and then passing it through our simulator (the \software{micrOMEGAs} package) to generate $X$. We begin sampling from $Y_0$ in the same way: we draw $\theta$ from the prior and plug it into our simulator to generate $X$. After this, however, we \textit{re-sample} $\theta ' \sim P(\theta)$ and use the new value in its place, i.e. $(X, \theta ') \sim Y_1$. 

We will use a neural network to model $r(X, \theta)$, which allows us to take advantage of the vast literature and resources available for deep learning applications. We will use an ensemble of multi-layer perceptrons (MLPs). Ensembling models and taking their average was shown in \cite{averting} to produce more accurate, conservative posteriors than would be obtained by simply training a single model.

Although not applicable for this work, it is worth pointing out that the data can be of virtually any format. For example, the authors of \cite{snre} have shown the ability of this algorithm to ingest 128$\times$128 pixel images of strongly-lensed galaxies and produce accurate posteriors of its Einstein radius.

\subsection{Sequential Neural Likelihood-to-Evidence Ratio Estimation}

If the prior is not carefully crafted to be a good approximation of the posterior to start with, we expect many of the samples to produce observables far from the experimental values of interest. These can be seen as wasted evaluations of the model, the very thing we are trying to limit. However, with a trained NRE, we have a better approximation of the posterior available to us. We therefore elect to iterate the training procedure outlined in the previous section, but replace the sampling of $\theta$ from the prior with a sampling from the intermediate posterior. The full algorithm is outlined in \AlgRef{alg:snre}.


Empirically, numerical experiments show that sequential versions of SBI tend to converge to the true posteriors with fewer simulator evaluations than their non-sequential counterparts \cite{sbi-benchmark}. Intuitively, this is because the approximate posterior will be most accurate in the densest regions of the parameter space \textit{as specified by the training data}. However, when we are sampling from the posterior, we are most interested in its regions of highest probability. Therefore, we can reduce the number of model evaluations required by iteratively closing in on the true posterior. We demonstrate this in Section \ref{sec:benefits} where we run trials to determine how efficient (the fraction of viable simulations) naive, NRE, and SNRE sampling, respectively, are. Our results are summarized in \FigRef{fig:post_eff} and \FigRef{fig:time_eff}.


Finally, it is important to note that either version of SBI is not guaranteed to converge to the true posterior and can often produce overconfident bounds on parameter space with exceedingly high computational budgets required to calibrate \cite{averting}. By using an ensemble of classifiers, which we do for all of our applications, we partially alleviate this problem.

\begin{algorithm}[H]
    \caption{Sequential Neural Likelihood-to-Evidence Ratio Estimation (SNRE)}\label{alg:snre}
    \begin{algorithmic}
    \State Let $p_r(\theta| X_0)$ be the posterior in round $r$. 
    \State Let $p_{r=0}(\theta | X_0) = P(\theta)$.
    \While{$r < R$}
    \State set prior to $p_r(\theta | X_0)$
        \While{$n \le N$}
            \State Sample $\theta_n \sim p_r(\theta | X_0)$
            \State Sample $X_i \sim P(X | \theta_n)$
            \State Add $\{\theta_n, X_n\}$ to $\mathcal{D}$
        \EndWhile
        \While{$d \le D$}
            \State Sample $\{\theta_A, X_A\}_d \sim \mathcal{D}$
            \State Sample $\{\theta_B, X_B\}_d \sim \mathcal{D}$
            \State Assign labels $y=1$ for $\{\theta_A, X_A\}_d$ and $\{\theta_B, X_B\}_d$
            \State Assign labels $y=0$ for $\{\theta_B, X_A\}_d$ and $\{\theta_A, X_B\}_d$
        \EndWhile
        \State (re-)train $r(X, \theta)$ to classify $\{\theta_A, X_A\}_D$, $\{\theta_B, X_B\}_D$, $\{\theta_B, X_A\}_D$, and $\{\theta_A, X_B\}_D$ by minimizing the binary cross-entropy loss.
        \State $p_r(\theta | X_0) = r(X, \theta) P(\theta)$
    \EndWhile
    \end{algorithmic}
\end{algorithm}

\section{Phenomenological Minimal Supersymmetric Model}
\label{sec:pmssm}
The fully general, unconstrained minimal supersymmetric extension to the Standard Model (MSSM) contains 105 free parameters~\cite{MARTIN_1998}, principally connected with the soft supersymmetry breaking pattern. By enforcing no flavor changing neutral currents, no new sources of CP violation, and first and second generation universality, these parameters are restricted to a 19-dimensional parameter space, known as the ``phenomenological'' MSSM (pMSSM)~\cite{pmssm}. Here, since we intend to illustrate the algorithms outlined above as applied to the pMSSM in connection with supersymmetric contributions to the anomalous magnetic moment of the muon, we entertain the possibility that the supersymmetric scalar particles associated with second generation leptons, smuons, be lighter than, and not as usually assumed degenerate with, their first generation counterparts, selectrons. Therefore, we relax the universality assumption in the slepton sector of the pMSSM, which adds two more parameters to the parameter space of interest. To simplify, and since they are largely irrelevant to quantities of interest, especially the muon anomalous magnetic moment, we assume all squark masses to be degenerate at a relatively large mass scale set to 4 TeV, large enough so that they have negligible effects in our calculations.

\begin{table}[ht]
    \begin{tabular}{@{}cccl@{}}
    \toprule
    Parameter    & Domain      & \phantom{ab} & Description                      \\ 
    \midrule
    $|\mu|$      & [100, 4000] & & Higgs mixing parameter          \\
    $|M_1|$      & [50, 1000]  & & Bino mass parameter              \\
    $|M_2|$      & [100, 4000] & & Wino mass parameter              \\
    $M_3$        & [400, 4000] & & Higgsino mass parameter          \\
    $M_{L_1}$    & [100, 4000] & & Left-handed selectron mass       \\
    $M_{L_2}$    & [100, 1000] & & Left-handed smuon mass           \\
    $M_{L_3}$    & [100, 4000] & & Left-handed stau mass            \\
    $M_{r_1}$    & [100, 4000] & & Right-handed selectron mass      \\
    $M_{r_2}$    & [100, 1000] & & Right-handed smuon mass          \\
    $M_{r_3}$    & [100, 4000] & & Right-handed stau mass           \\
    $M_A$        & [400, 4000] & & Pseudoscalar Higgs mass          \\
    $\tan \beta$ & [1, 60]     & & Ratio of Higgs VEVs $v_{u}/v_{d}$              \\
    $|A_t|$      & [0, 4000]   & & Trilinear Higgs-stop coupling    \\
    $|A_b|$      & [0, 4000]   & & Trilinear Higgs-sbottom coupling \\
    $|A_{\tau}|$ & [0, 4000]   & & Trilinear Higgs-stau coupling    \\ 
    \bottomrule
    \end{tabular}
    \caption{Domains used for the pMSSM soft supersymmetry breaking parameters. All parameters (aside from $\tan\beta$) are in units of GeV. We set all squark mass parameters to 4 TeV.}
    \label{tab:pmssm_parameters} 
\end{table}


To utilize the SBI framework, one must have a forward model of the likelihood, i.e., a function which takes in a set of parameters, $\theta$, and produces an observable, $X$, from some underlying distribution. The underlying distribution need not be known explicitly, for our goal is to learn it from the samples. Additionally, we must provide a prior distribution of the parameters. For our applications, we will use a uniform prior for the parameters, with bounds chosen to be consistent with the literature and listed in \TabRef{tab:pmssm_parameters}.

In the following applications, we aim to produce posterior distributions conditioned on a combination of observables calculated by the \software{micrOMEGAs} package \cite{micromegas}. Although not arbitrarily accurate, \software{micrOMEGAs} produces a {\em deterministic} output, so we assume Gaussian error bars on these calculations. Thus, our likelihood is given by $\mathcal{N}\left(X(\theta), \sigma_X \right)$, where $\mathcal{N}$ is the normal distribution and $X(\theta) = (\Omega_{\textrm{DM}}(\theta) , a_\mu(\theta) , m_h(\theta), p_{\textrm{Xenon1T}})$ is the vector of calculated observables from \software{micrOMEGAs}. The variance of the distribution is $\sigma_X^2 = \sigma_{\textrm{exp}}^2 + \sigma_{\textrm{th}}^2$  with theoretical Gaussian uncertainties are: 
\begin{align}
\sigma_X 
&= (\sigma_\Omega, \sigma_{\asusy}, \sigma_{m_h}, \sigma_{p_{\textrm{Xenon1T}}}) \\
&= (0.01, 65\times 10^{-11}, 1.0 \textrm{\ GeV}, 10^{-5}).\notag
\end{align}
pMSSM parameters, $\theta$ are sampled from a uniform prior, with bounds shown in table \ref{tab:pmssm_parameters}. When sampling from our posterior, we set the observed values of $X$ to their experimental results~\cite{Planck, higgs1, higgs2, gmuon1, gmuon2} 
\begin{align}
X_0 
&= (\rdchi, m_{h}, \asusy, p_{\mathrm{Xenon1T}})\\ 
&= (0.12, 125\ \textrm{GeV}, 251\times 10^{-11}, 0.0455).\notag
\end{align}
Here we are assuming that the discrepancy between the SM and experimentally observed anomalous magnetic moment of the muon is solely due to supersymmetric contributions $\asusy = a^{\mathrm{exp.}}_{\mu}-a_{\mu}^{\mathrm{SM}}$. As Xenon1T has not yet observed WIMP dark matter, we set its p-value equal to $\textrm{CDF}_{\mathcal{N}}(2\sigma)$ and treat all p-values greater than this value identically via the pre-processing described in \EqnRef{eq:preprocessing_omega}-\EqnRef{eq:preprocessing_xenon}.

\section{Benefits from Sequential Training}
\label{sec:benefits}
A key question in algorithm optimization is to seek the minimization of the time required to obtain $N$ samples from a distribution. Here, we can approximate the amount of time spent computing, $T$, via 
\begin{equation}
    \label{eq:compute_time}
    T = N s + R t + (R - 1) v + w
\end{equation}
where $N$ is the total number of \software{micrOMEGAs} calculations performed, $s$ is the time per \software{micrOMEGAs} evaluation, $R$ is the number of sequential rounds of training, $t$ is the time to train the network in each sequential round, $v$ is the amount of time to sample from the intermediate posteriors during SNRE training, and $w$ is the amount of time to sample from the final SNRE posterior. In this application, we limit $R, t, v, w$, so that the majority of the time is spent evaluating \software{micrOMEGAs}, i.e. $T \approx N s$. $N$ is often chosen to be large enough so that the number of samples surviving the constraints is larger than some number, $M$. We can write 
\begin{equation}
    N = M/\varepsilon + D,
\end{equation}
where $\varepsilon$ is the efficiency of our \software{micrOMEGAs} samples, and $D$ is the number of samples used to train our models (for traditional methods, $D = 0$). If the efficiency is sufficiently small, the number of required samples becomes prohibitively large, which is the typical case when sampling $\theta$ from the prior. The goal of SBI is to minimize $N$ by maximizing $\varepsilon$. We illustrate this effect by calculating the efficiency of over a range of training sample sizes and SNRE training rounds.

\subsection{Setup}

The benefits of the SBI framework become clear when looking at the efficiency of sampling, i.e., the fraction of samples that satisfy the given constraints of the problem. To illustrate this, we perform the pMSSM posterior approximation exercise while enforcing observables to be within $3$-sigma of their central values:
\begin{equation}\label{eq:filter}
\begin{split}
    0.09 < \  &\Omega_\chi h^2 < 0.15 \\
    122\  \textrm{GeV} < \  &m_h < 128\  \textrm{GeV} \\
    56 \times 10^{-11} < \  &\asusy < 445 \times 10^{-11} \\
    0.0455 < \  &p_{\textrm{Xenon1T}}
\end{split}
\end{equation}
where $\asusy = (g^{\mathrm{SUSY}}_{\mu}-2)/2$ is the supersymmetric contribution to the muon anomalous magnetic moment and $p_{\textrm{Xenon1T}}$ is the p-value of the model as determined by Xenon1T. 

We choose to run the SBI pipeline on three different choices of $R$. Specifically we set $R=1$ for NRE, $R=3$ for SNRE$_3$, and $R=5$ for SNRE$_5$. 

We compare each choice of $R$ with the ``prior'' baseline, which calculates observables using samples from the prior, and then subjects them to the same filtering as shown in \EqnRef{eq:filter}. We emphasize that samples from the prior do not correspond to samples from the posterior, and we only use the samples as a point of comparison.

We allot a budget of 150,000 \software{micrOMEGAs} calculations and split them into training sets and inference sets. We use the training sets to train the networks, and the inference sets are the evaluations from which our final samples come. Ideally, the training sets will be as small as possible to maximize the number of posterior samples in our inference set. However, increasing the size of the training set could increase the efficiency of posterior samples, thus enabling equal performance with the weaker algorithm producing a larger inference set. For each method, we run a trial on a training set of sizes 10,000, 31,000, and 100,000, with the remaining calculations left for inference. For SNRE$_R$, we distribute the calculations equally amongst the $R$ rounds.

\subsection{Results}

\begin{figure}
    \begin{centering}
    \includegraphics[width=0.9\columnwidth]{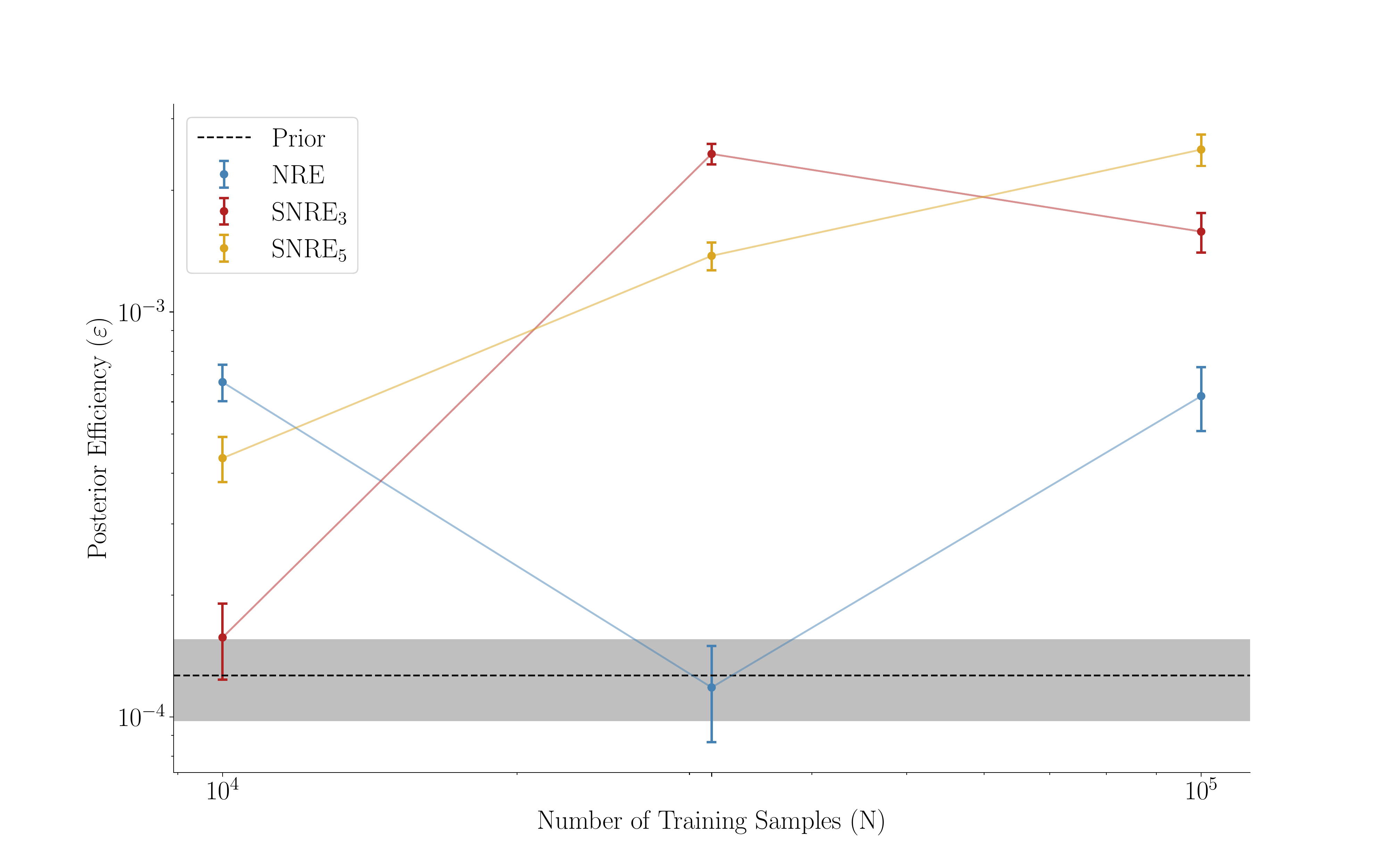}
    \caption{The fraction of points sampled from the posterior that lie within the ranges specified in \EqnRef{eq:filter}. The grey band indicates the efficiency of sampling directly from the prior for comparison with standard MCMC based approaches.}
    \label{fig:post_eff}
    \end{centering}
\end{figure}


\begin{figure}
    \begin{centering}
    \includegraphics[width=0.9\columnwidth]{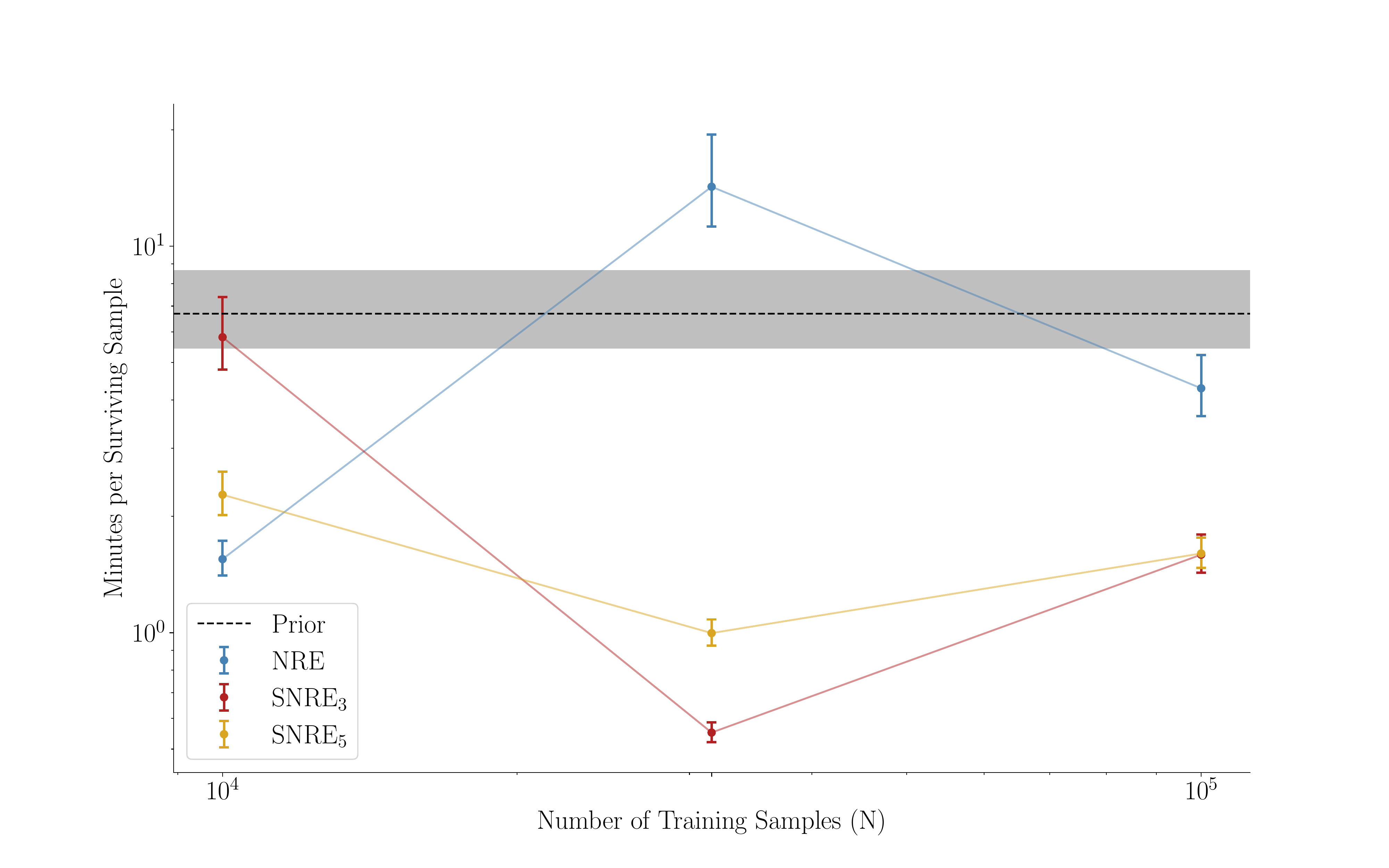}
    \caption{The expected amount of time (minutes) needed to obtain one sample in the ranges specified in \EqnRef{eq:filter}.}
    \label{fig:time_eff}
    \end{centering}
\end{figure}

\begin{figure*}
    \begin{centering}
    \includegraphics[width=0.9\textwidth]{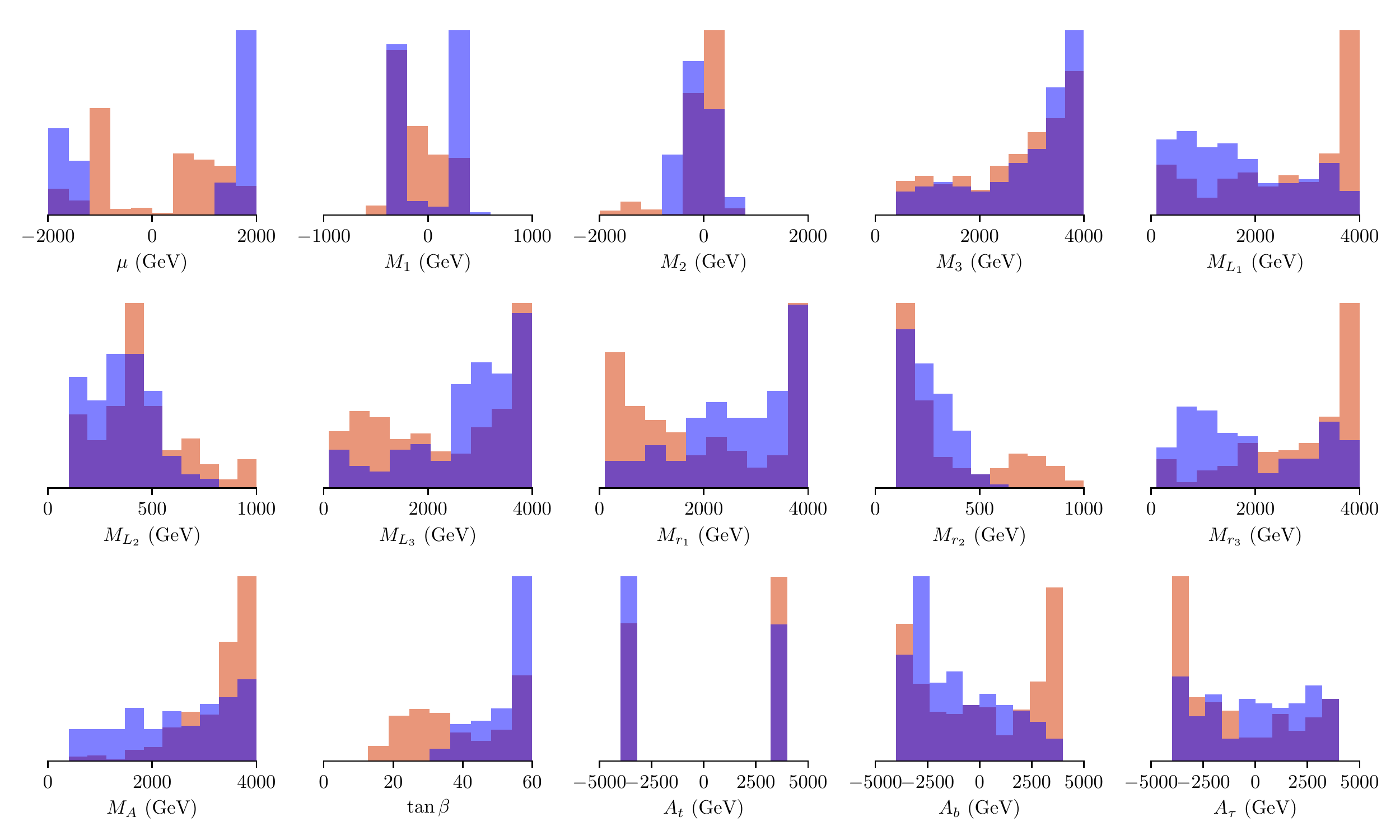}
    \caption{Samples from the posterior which lie within the experimental constraints specified in \EqnRef{eq:filter} obtained by running SNRE with hyperparameters listed in \TabRef{tab:hyperparameters} are shown in salmon. We note the overdensity of small smuon masses which affect $a_\mu$ and $\Omega_\chi$ by coannhiliations. See \FigRef{fig:filtered_masses} and \FigRef{fig:filtered_gaugino} for corner plots with subsets of these parameters. Shown in blue are the filtered samples from the posterior for the non-sequential NRE, again trained using the hyperparameters in \TabRef{tab:hyperparameters}. We see the SNRE algorithm produces similar coverage for most parameters while providing greater sampling efficiency.}
    \label{fig:1d_hists}
    \end{centering}
\end{figure*}

The results demonstrate several notable features. First, in \FigRef{fig:post_eff} we observe the expected trend amongst the sequential algorithms; as the number of points used to train increases, the number of points sampled from the approximate posteriors which lie within the ranges specified in \EqnRef{eq:filter} also increases. However, this is not true for NRE, which performed worse for the medium-sized training set. The absence of this scaling for NRE is likely due to overfitting, which we did not prevent by early-stopping. Finally, as the efficiency is highest for SNRE$_5$, it is the preferred model in the event that additional computational resources become available after the training phase, e.g., in future follow-up analyses.

We determine the average amount of time from start to finish required to generate a valid sample from the final posterior and display the results in \FigRef{fig:time_eff}. Although slower overall, we determine that the efficiency of SNRE$_R$ gives it a factor of up to 3 times faster sampling than NRE, even when the inference set is significantly smaller. Our results suggest sequential algorithms are consistently the most efficient when they use $\mathcal{O}(10\%)$ of the allotted calculations for training.

\section{Application to pMSSM}
\label{sec:sbi-pmssm}

\subsection{SBI Setup}

We now examine the posteriors and experimental observables produced by an application to the pMSSM. In this search, we are interested in finding the regions of parameter space most likely to contain the correct relic density, Higgs mass, provide a good fit to the observed anomalous magnetic moment of the muon, and are not excluded by Xenon1T.

Inspired by the results of section~\ref{sec:benefits}, we run SNRE for a total of 10 rounds, each with 20,000 new samples calculated with \software{micrOMEGAs}. Next, we assemble the likelihood-to-evidence approximators with an ensemble of 5 binary classifiers whose inputs are a concatenation of $X$ and $\theta$ and whose outputs are the logit of the classification probability. Each network consists of three hidden layers, each consisting of 256 units. Next, we train networks to minimize the binary cross-entropy loss between samples from the likelihood, $p(X | \theta)$, and the joint probability, $p(X)p(\theta)$. Finally, we average the ensemble outputs during inference time to produce an approximate likelihood-to-evidence ratio of a given sample. When we sample with HMC during intermediate training steps, we use 20 chains with 2000 warmup steps and 32 chains with 2000 warmup steps during the final inference time.

\begin{figure}
    \begin{centering}
    \includegraphics[width=0.9\columnwidth]{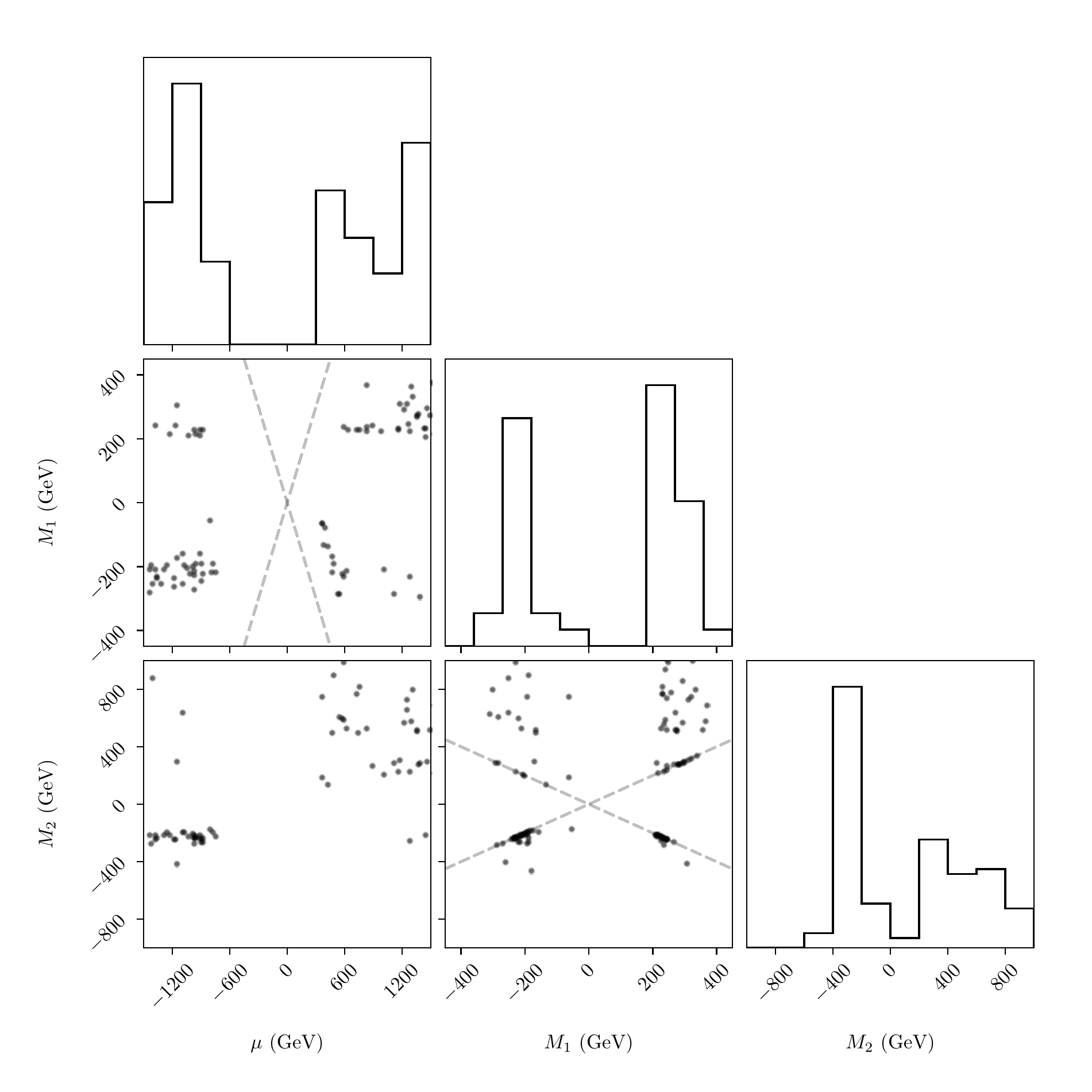}
    \caption{Corner plots of $\mu$ and gaugino mass parameters created from the same samples as \FigRef{fig:1d_hists}. Dotted lines correspond to $\mu = M_1$ and $M_2-2 = M_1$, respectively. We see a strong bias towards bino-like and wino-like LSPs. Direct detection constraints force $\mu$ towards larger values of its range. Additionally, we see a dependence on the relative signs of $\mu$ and $M_2$.}
    \label{fig:filtered_gaugino}
    \end{centering}
\end{figure}

\begin{figure}
    \begin{centering}
    \includegraphics[width=0.9\columnwidth]{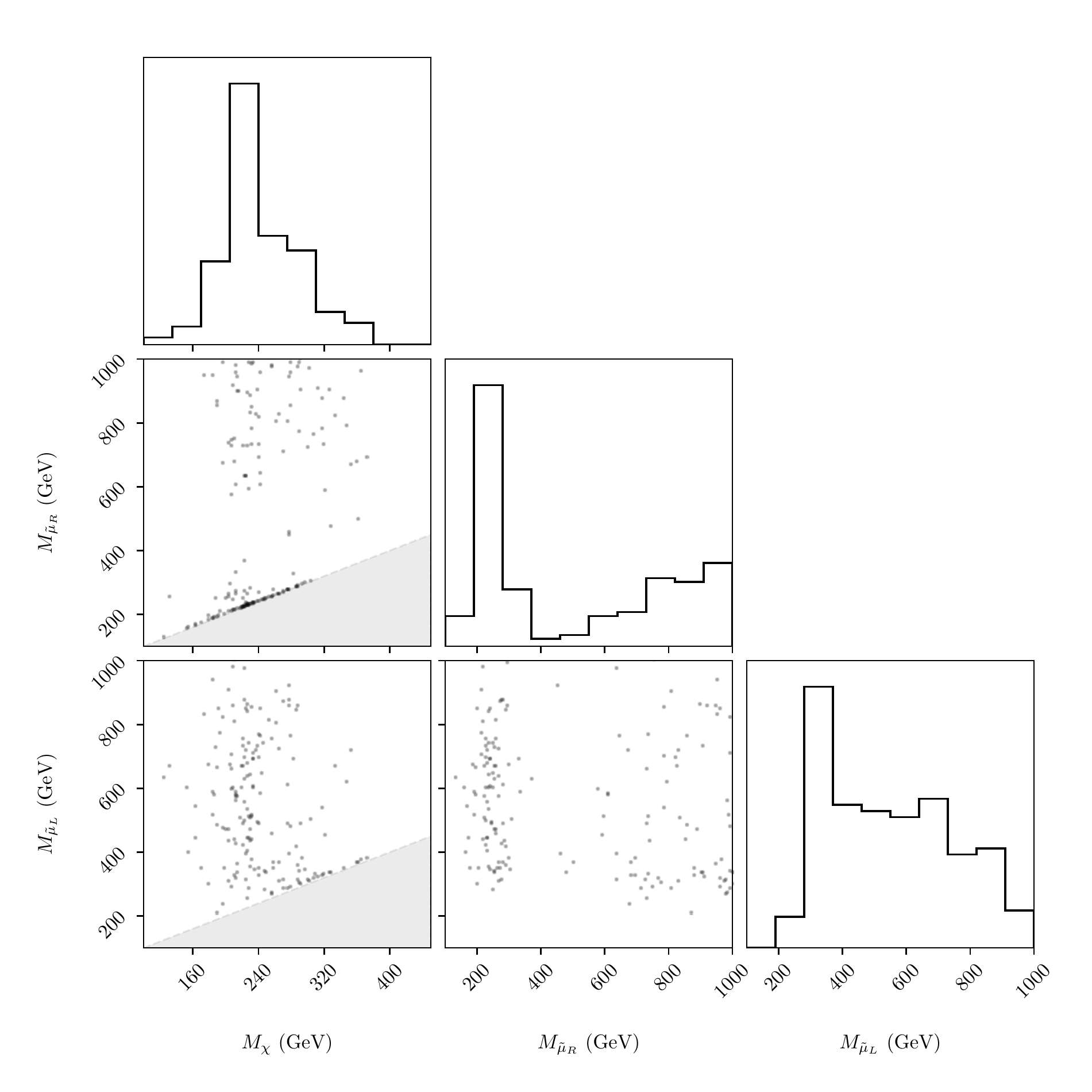}
    \caption{Corner plots of smuon and dark matter masses calculated from the same samples as \FigRef{fig:1d_hists}. Shaded regions do not have a neutralino LSP which, although not explicitly excluded, are not experimentally viable. We note the large concentration of light right-handed smuons.}
    \label{fig:filtered_masses}
    \end{centering}
\end{figure}

\begin{figure}
    \begin{centering}
    \includegraphics[width=0.9\columnwidth]{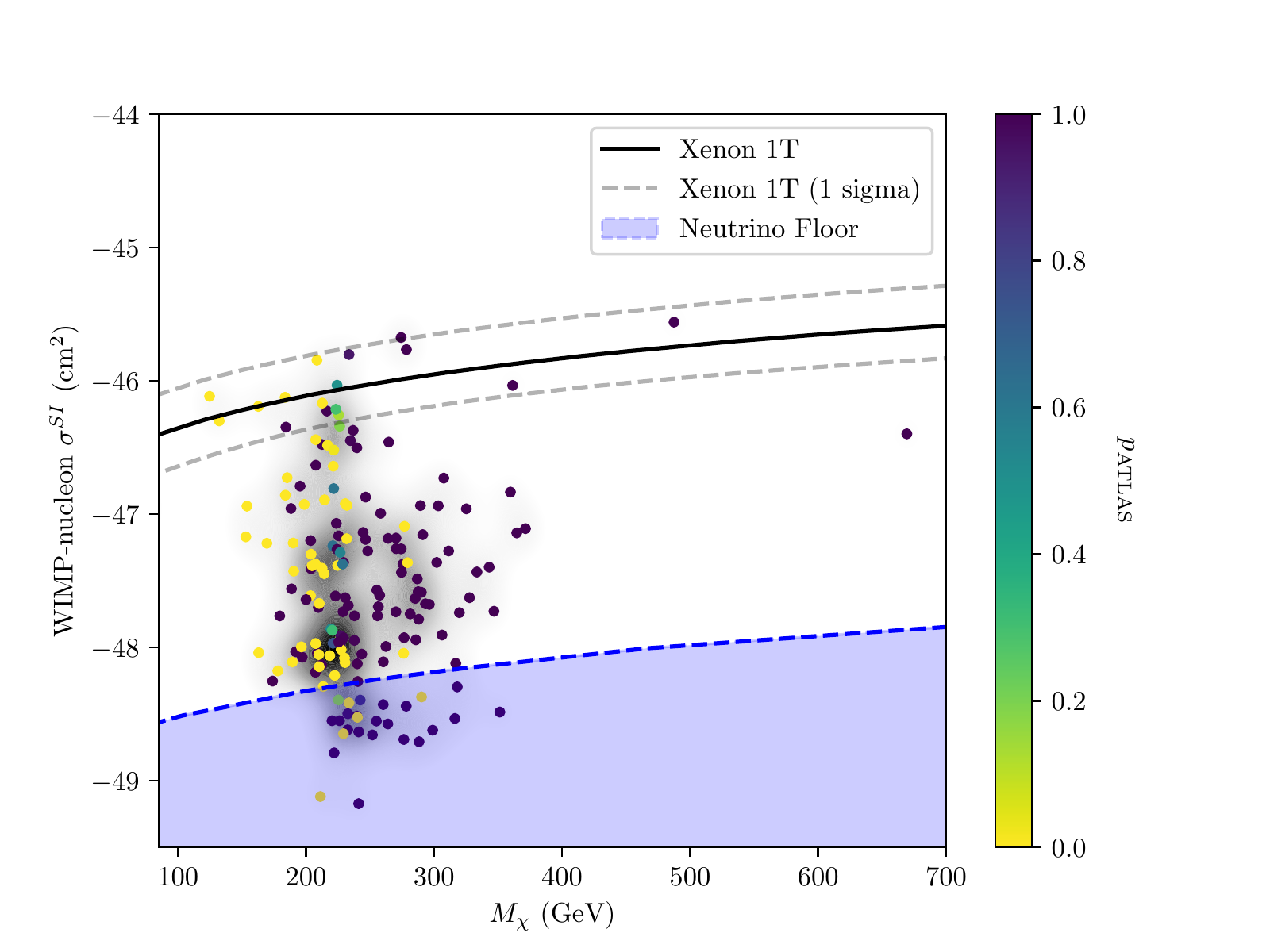}
    \caption{Samples from the approximate posterior which lie within the experimental constraints specified in \EqnRef{eq:filter}. The color of each point corresponds to the p-value from compressed spectra constraints reported by ATLAS. The background is shaded according to a gaussian kernel density estimate in order to visualize the concentration of points on these axes. We note the ability of ATLAS to probe models which lie in the neutrino floor.} 
    \label{fig:filtered_DD}
    \end{centering}
\end{figure}

\begin{figure}
    \begin{centering}
    \includegraphics[width=0.9\columnwidth]{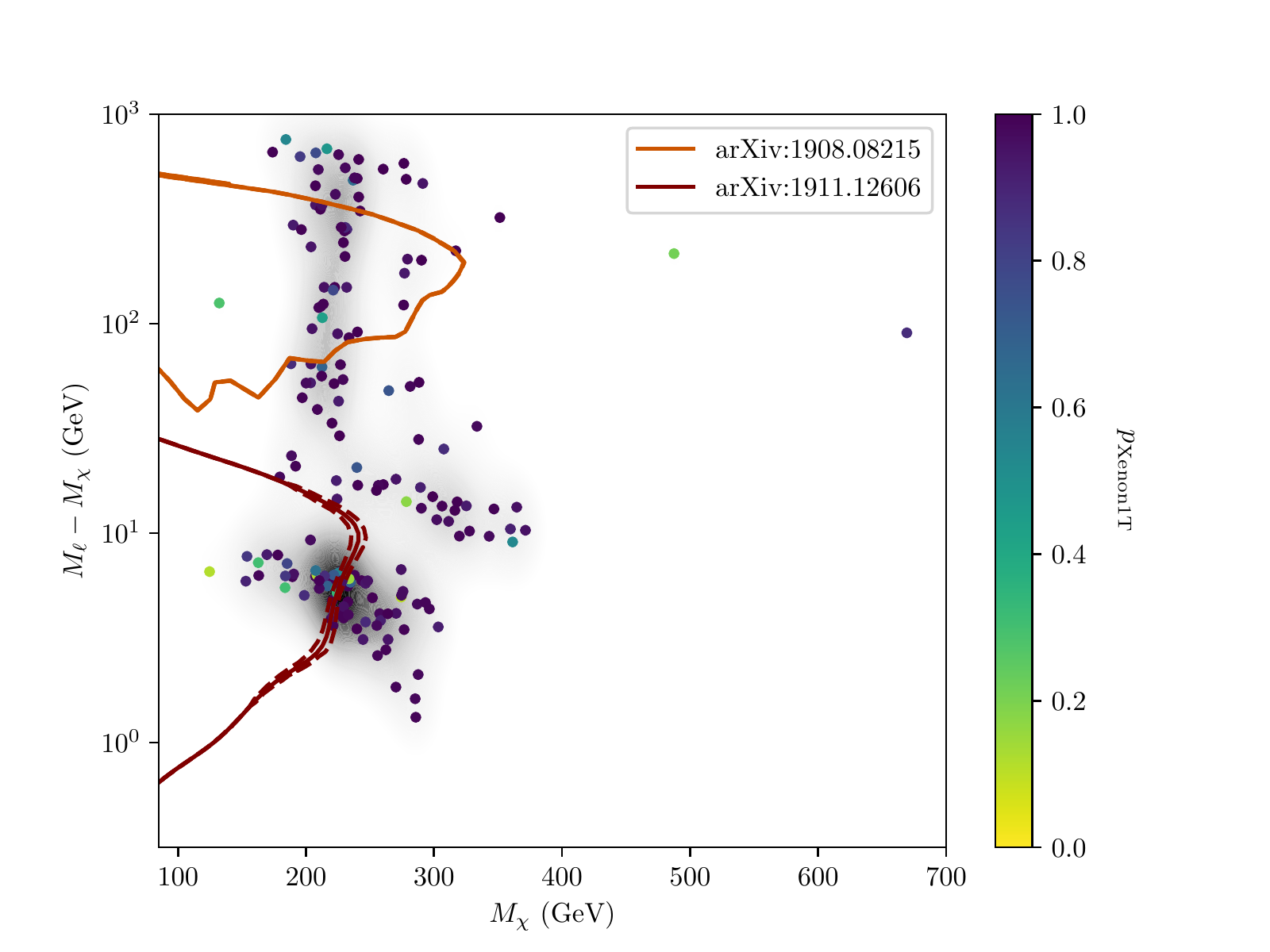}
    \caption{Same as \FigRef{fig:filtered_DD} but plotted on ATLAS constraints with Xenon1T p-values, as calculated by micrOMEGAS. The mass splitting is taken with respect to the lightest slepton mass, $M_\ell$. Dotted lines correspond to $\pm 1\sigma$ constraints. We expect future analyses to constrain much of the viable parameter space with small mass splittings.}
    \label{fig:filtered_LHC}
    \end{centering}
\end{figure}

\subsection{Results}

In \FigRef{fig:1d_hists}, we show samples from the approximate posterior as one-dimensional histograms. The multi-modal and irregularly shaped distribution of points demonstrates the ability of flexible neural methods to hone in on these regions. The results exhibit several features previously found in the literature despite not being enforced a priori in our search (outside of choosing bounds on the prior.) Additionally, we plot two-dimensional histograms for a subset of the dimensions in \FigRef{fig:filtered_gaugino} and \FigRef{fig:filtered_masses}.

These results demonstrate, for example, the learned dependence on the relative signs of $\mu$ and the gaugino parameters. In addition, we find that most models have a bino-like LSP with light smuons whose masses are near the LSP mass. Physically, these correspond to the bino coannihilation with smuons driving thermal relic abundance. One expects these results as bino-like LSPs have smaller WIMP-nucleon cross sections, and lighter sleptons contribute more to $a_\mu$, while also suppressing the relic density. We note the generally larger values of $M_{L_2}$ compared to $M_{r_2}$. This trend is mostly due to the sneutrino requiring a larger value of $M_{L_2}$ because its mass is typically lighter than the left-handed smuon. The right-handed smuon mass, on the other hand, is generally equal to $M_{r_2}$.

Of the 96,000 samples generated from the approximated posterior, 179 survive the cuts laid out in \EqnRef{eq:filter}, resulting in an efficiency of 0.00186, similar to the results shown in \FigRef{fig:post_eff}. The surviving points are shown on top of Xenon1T and ATLAS  \cite{MET_search, compressed_search} constraints in \FigRef{fig:filtered_DD} and \FigRef{fig:filtered_LHC}, respectively.  Interestingly, our results indicate that the experiments are complementary: regions of low direct detection constraining power exist in high ATLAS constraining power regions, and vice versa. We expect future analyses to probe the majority of the surviving parameter space.

\section{Conclusions}
\label{sec:conclusion}
We introduced a simulation-based inference framework capable of analyzing high-dimensional parameter spaces and applied the framework to supersymmetric extensions to the Standard Model. We showed how equivalent amounts of computational time result in orders of magnitude more pMSSM models of experimental interest than naive sampling from the prior. Additionally, we demonstrated how sequential sampling methods are even more efficient than their non-sequential counterparts by performing a light scan over hyperparameters.

Finally, we applied the framework to sample from regions of the pMSSM parameter space yielding the correct thermal dark matter relic density, Higgs mass, and the muon's anomalous magnetic moment, which direct detection experiments have not yet excluded. We found that the most likely regions of this space are tantalizingly close to current ATLAS bounds and are likely to be covered in future direct dark matter detection experiments. 

Future exercises similar to the present one could make use of calibration methods presented in Ref.~\cite{calibrated-classifiers} to avoid overly-confident constraints on the approximated posterior and therefore avoid excluding potentially viable regions of parameter space. Additionally, one can consider higher-dimensional observables. Rather than summaries of observed weak-scale quantities, it is potentially of interest to operate directly on the raw data obtained from experiments.

\begin{acknowledgments}
The authors would like to thank Michael Hance, Giordon Stark, Jeff Shahinian, and Michael Holzbock. This work is partly supported by the U.S.\ Department of Energy grant number de-sc0010107 and the National Science Foundation Graduate Research Fellowship under Grant No. DGE-1842400 to NS.
\end{acknowledgments}

\appendix

\section{Training details}
Prior to entering neural network, is data are pre-processed to restrict the dynamic range of the inputs. For the relic density, we enforce the value to be greater than $10^{-4}$, convert to log-space and rescale:
\begin{align}\label{eq:preprocessing_omega}
\Omega_\chi h^{2} &\rightarrow 10 \times \log_{10}\left(\textrm{clip}(\Omega_\chi h^{2}, 10^{-4}, \infty)\right)
\end{align}
The Higgs mass is restricted to be larger than $118$ GeV and made $\order{1}$ by subtracting off a value close to its know mass $123.864$ GeV and rescaling to be $\order{1}$
\begin{align}\label{eq:preprocessing_higgs}
m_h &\to \frac{\mathrm{clip}(m_h, 118, \infty) - 123.864}{2.2839}
\end{align}
The SUSY contribution to the muon anomalous magnetic moment is restricted to be within $10^{-11}$ and $10^{-8}$, converted to log-space and made $\order{1}$
\begin{align}\label{eq:preprocessing_asusy}
\asusy &\to \log_{10}\mathrm{clip}(a_\mu, 10^{-11}, 10^{-8}) + 9.5
\end{align}
Lastly, we restrict the p-value from Xenon1T to be positive and less than $0.0455$. We then rescale to be $\order{1}$:
\begin{align}\label{eq:preprocessing_xenon}
p_{\textrm{Xenon1T}} &\to  10 \times \mathrm{clip}(p_{\textrm{Xenon1T}}, 0, 0.0455)
\end{align}
In the above, $\mathrm{clip}(x, x_{\mathrm{min}}, x_{\mathrm{max}})$ is given by:
\begin{equation}
  \mathrm{clip}(x, x_{\mathrm{min}}, x_{\mathrm{max}}) =
  \begin{cases}
       x_{\mathrm{min}} & \text{if $x < x_{\mathrm{min}}$} \\
       x_{\mathrm{max}} & \text{if $x > x_{\mathrm{max}}$} \\
       x & \text{otherwise}
  \end{cases}
\end{equation}
The hyperparameters used in the networks are listed in \TabRef{tab:hyperparameters}.

\begin{table}[!ht]
\begin{tabular}{@{}llr@{}}
\toprule
    Hyperparameter      & & Value            \\ \midrule
    Hidden layers       & & 3                \\
    Hidden layer size   & & 256              \\
    Ensemble size       & & 5                \\
    Learning rate       & & 3 $\times 10^{-4}$ \\
    Weight decay        & & 0                \\
    Max gradient norm   & & $10^{-3}$          \\
    Validation interval & & 50              \\
    Patience            & & 200              \\
    Batch size          & & 256              \\
    Training split      & & 0.95             \\ \bottomrule
\end{tabular}
\caption{Hyperparameters common to all (S)NRE algorithms tested in all applications.}
\label{tab:hyperparameters}
\end{table}

We used the \software{SaxBI} package \cite{Tamanas_SaxBI_2022} with \software{JAX} \cite{jax} and \software{Flax} \cite{flax} backends to initialize and fit the models to perform approximate Bayesian inference. We used the \software{optax}-repository's~\cite{optax} implementation of the Adaptive Moment Estimation (Adam) algorithm \cite{adam} to optimize our models. We used \software{numpyro}'s implementation of Hamiltonian Monte Carlo (HMC) \cite{numpyro1} \cite{numpyro2} to sample from the posterior distribution of the model parameters. We run our simulations using \software{microMEGAS} v5.2.13 \cite{micromegas} with \software{SOFTSUSY} v.4.1.7 \cite{softsusy} backend. Our entire codebase is open-source and can be found here: \url{https://www.github.com/jtamanas/MSSM} \faGithub. All computations were run on a machine with an AMD Ryzen 7 3700X processor, 16 GB of RAM, and no GPU.\\




\bibliography{main}

\end{document}